\documentstyle[12pt,iopconf]{article}
\begin{document}
\title{Acousto-optic effects of surface acoustic waves in semiconductor quantum well structures}

\author{C. Rocke\dag\footnote{E-mail: Carsten.Rocke@physik.uni-muenchen.de.}, 
 A. Wixforth\dag, and J.P. Kotthaus\dag\\W. Klein\ddag\ , H. B\"ohm\ddag\ , and G. Weimann\S}

\affil{\dag\ LMU M\"unchen, Geschw. Scholl Platz 1, D-80539 M\"unchen, Germany}

\affil{\ddag\ Walter-Schottky-Institut der TUM, D-85748 Garching, Germany}

\affil{\S Fraunhofer Institut IAF, Tullastrasse, D-79108 Freiburg, Germany}

\beginabstract

We report on recent experiments investigating the modification of the inter-band optical response of a piezoelectric semiconductor quantum well structure under the influence of intense short period surface acoustic waves. Experimentally, we study the photoluminescence (PL) of an undoped strained InGaAs/GaAs quantum well on which surface acoustic waves are propagated in the GHz regime. We observe a pronounced influence on the PL, both in intensity as well as in energetic position: Above a critical acoustic power density and corresponding lateral piezoelectric field strength, we observe a quenching of the excitons resulting in a strong decrease of the PL intensity. Using two transducers in a cavity resonator geometry, we can create a standing surface acoustic wave and hence control the nature and efficiency of the acoustic transport of the photoexcited electrons and holes.

\endabstract

\section{Introduction}

Surface acoustic wave (SAW) studies of low dimensional electron systems confined in piezoelectric semiconductors have proven to be a powerful tool for the investigation of the dynamical conductivity of the system \cite{achim}. SAW frequencies up to several GHz and correspondingly short wavelengths in the submicron regime are nowadays accessible. The interaction between the mobile carriers as confined in one dimension in a semiconductor heterojunction are found to be dominated by the piezoelectric fields accompanying the SAW at the speed of sound.

\begin{figure}
\vspace*{7.8cm}
\hspace*{1cm}
\special{epsf="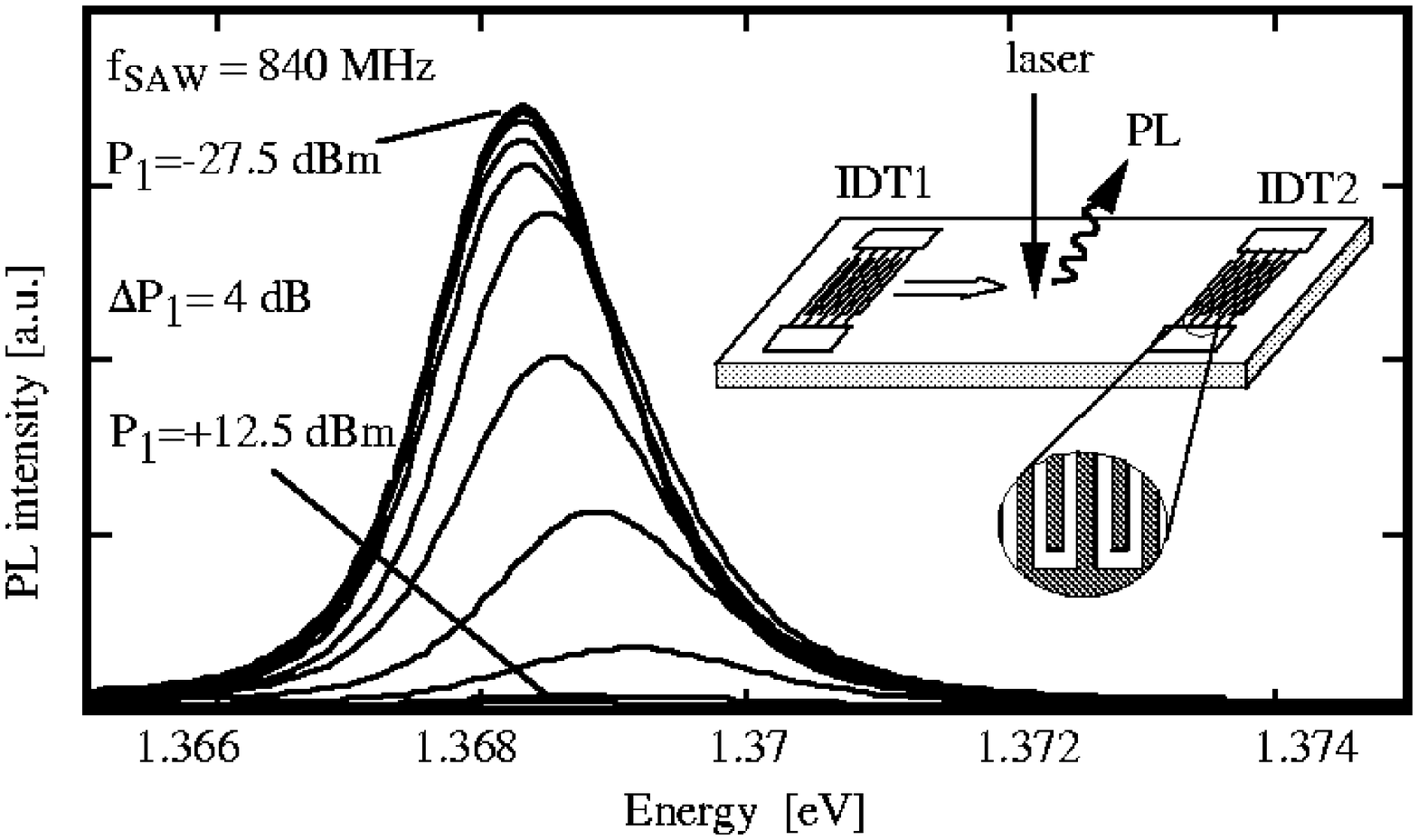"}
\caption{Photoluminescence spectra of a single 10nm wide InGaAs/GaAs QW structure for different applied SAW powers. The laser excitation intensity was 10mW/cm$^2$ at an energy of 1.49 eV.}
\end{figure}

The influence of the SAW on the optical properties of the semiconductor heterostructures on the other hand have so far received only limited attention. Acousto-optic modulators utilizing surface acoustic waves in III-V semiconductor multiple quantum wells (MQW's) have been suggested for optical signal processing applications such as electrically addressable spatial light modulators \cite{modulator}. There, the electric fields generated by the SAW in the piezoelectric material modulate the optical properties via the quantum confined Stark effect (QCSE). In addition, deformation potential coupling  of the SAW-related strain fields may have pronounced effects on the optical properties as has recently been shown theoretically by Smith et al. \cite{smith}.

\section{Experimental results}

We investigate the low-temperature photoluminescence of a single undoped 10nm wide InGaAs/GaAs QW under the influence of intense surface sound waves in the GHz regime. On both ends of the sample several lithographically defined interdigital transducers (IDT) operating at different frequencies are fabricated by successively evaporating NiCr and Al directly on the surface. The interdigital electrode spacing establishes the fundamental acoustic wavelength $\lambda_0$ and frequency $f_0$ by $f_0=v/\lambda_0$, where $v$ is the sound velocity. The SAW is launched by the application of a HF signal at the fundamental or harmonic frequency to one of these IDT. A tunable Titanium-Sapphire laser is used for the optical excitation and the PL signal is detected using a triple grating spectrometer.

Figure 1 shows PL spectra at T=4K for several SAW powers. As the acoustic intensity is increased, the excitonic transition is shifted to slightly higher energies and finally quenched. We attribute this effect to the vertical $E_v$ and lateral $E_l$ electric fields generated by the SAW. For the highest acoustic power of  $P_1 = 13.5$ dBm (22.4 mW) electric fields as high as $E_l=8$ kV/cm and $E_v=10$ kV/cm are achieved, resulting in a carrier escape via tunneling and field-ionization of the excitons. Recent related studies using statically imposed lateral electric fields on the same wafer also showed a decrease of the PL-intensity reflecting the field ionization and a similar blue shift for comparable electric fields. \cite{andy}.

\begin{figure}[tb]
\vspace*{9.6cm}
\hspace*{1cm}
\special{epsf="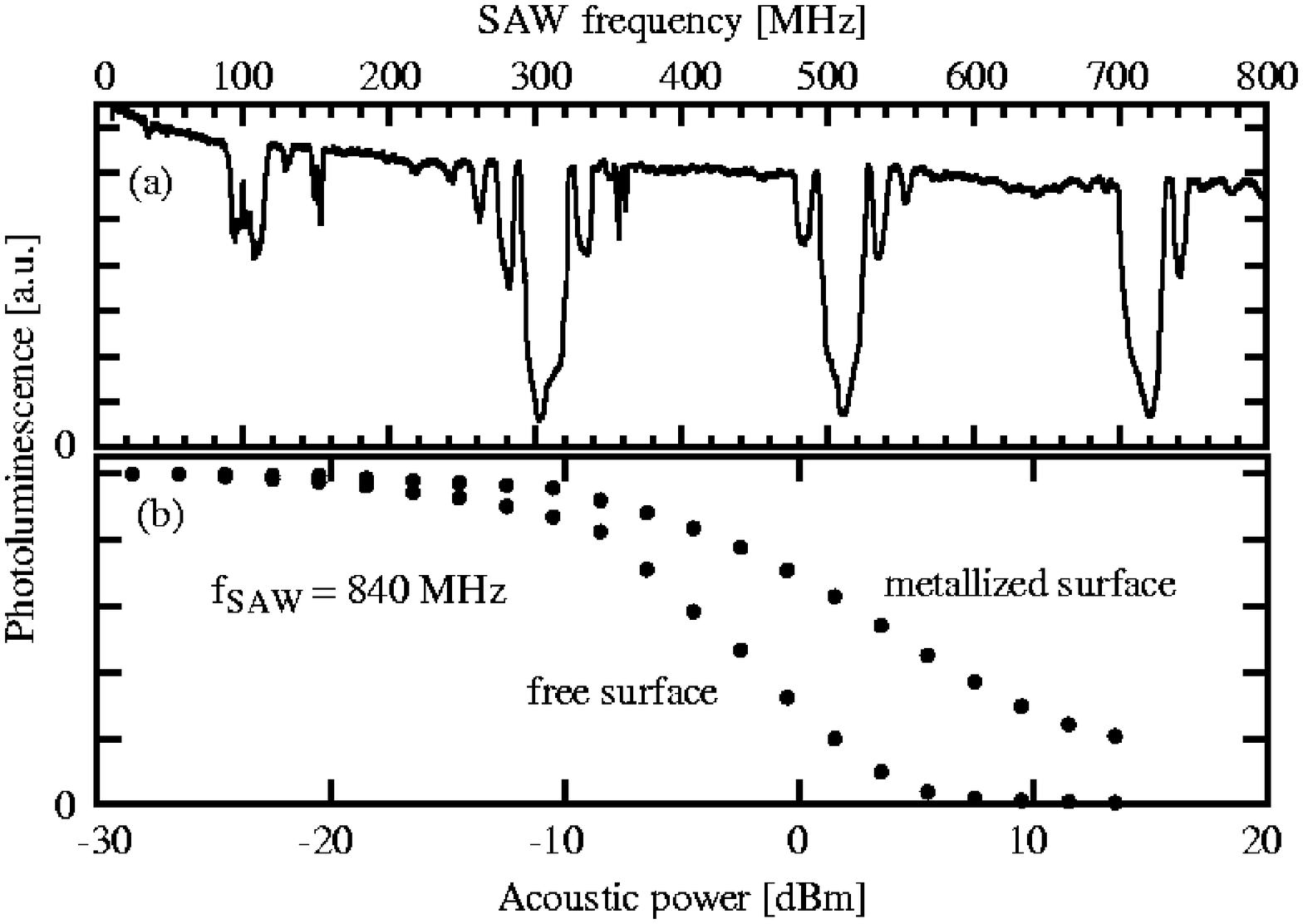"}
\caption{(a) Photoluminescence intensity as a function of the SAW frequency applied to a 4-finger per electrode IDT. The 4 primary harmonic resonances manifest themselves in pronounced quenching of the excitonic transition and a decrease of the PL intensity.
(b) PL intensity on a free and a metallized part of the surface as a function of the applied acoustic power. The screening of the lateral piezoelectric fields by the semitransparent metallization results in a partial recovery of the PL signal.}
\end{figure}

In Fig. 2(a) we plot the PL intensity as a function of the SAW frequency using a multi-frequency IDT on a different sample. As the frequency is varied the decrease of the PL signal directly maps the frequency response of the IDT, thus clearly indicating the acoustic nature of the interaction. To distinguish between effects related to the different piezoelectric field polarizations, we used a thin semitransparent metallization on part of the sample which readily screens the lateral fields imposed by the SAW. This screening results in a recovery of the PL in comparison to the free surface as depicted in Figure 2(b). The influence of the vertical piezoelectric field and deformation potential coupling can thus be studied independently. A dramatic increase in linewidth for the highest acoustic powers used is also observed and attributed to the QCSE and deformation potential coupling which both influence the transition energies.

For the highest acoustic powers, the potential modulation caused by the lateral piezoelectric fields is strong enough to localize the spatially separated electrons and holes in the potential extrema of the conduction and valence band, respectively. As this spatial separation is of the order of $\lambda_0/2=1.7 \mu$m, the strongly diminished wavefunction overlap results in a pronounced reduction of the radiative transition probability. Figure 3 shows the PL intensity as a function of the acoustic power $P_2$ for a constant power $P_1 = 5.5$ dBm. As $P_2$ is increased, the initially unidirectional wave is converted into a standing wave. The spatial separation of electrons and holes is thereby lifted which results in a recovered and increased PL signal. For $P_2 \gg  P_1$ the SAW becomes travelling again with reversed propagation direction and larger amplitude. Hence, it efficiently transports the localized electrons and holes to the other side of the detection area which is indicated by the strong decrease in PL intensity.

\begin{figure}[tb]
\vspace*{9.5cm}
\hspace*{0.5cm}
\special{epsf="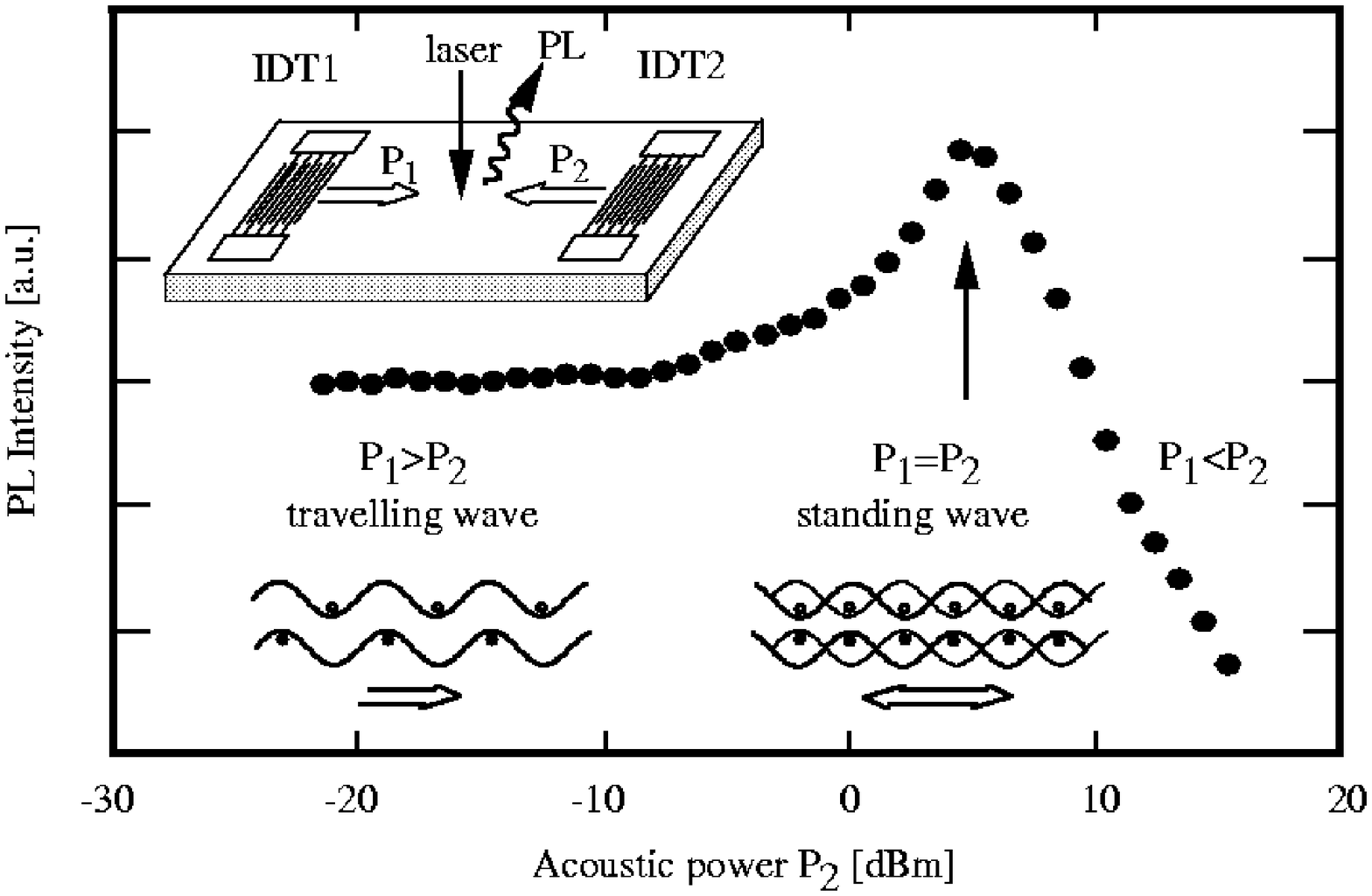"}
\caption{Photoluminescence intensity as a function of the acoustic power from IDT2 for a constant acoustic power $P_1 = +5.5$ dBm. The SAW frequency was 840 MHz.}
\end{figure}

We gratefully acknowledge useful discussions with A.  Govorov and experimental help of M. Rotter and the financial support by the Deutsche Forschungsgemeinschaft (SFB 348) and the Bayerische Forschungsverbund FOROPTO.

\end{document}